\documentclass[preprint,showpacs,superscriptaddress,floatfix,prc]{revtex4}
\usepackage{epsfig}
\usepackage{epsf}
\usepackage{psfig}
 
\usepackage{longtable}
\usepackage{graphicx}
\usepackage{bm}
 
\begin{document} 
 
 

\title{Two-phonon $1^-$ state in $^{112}$Sn observed in resonant photon scattering} 
 
\author{I.~Pysmenetska}
  \affiliation{Institut f\"ur Kernphysik,  
               Technische Universit\"at Darmstadt, 
               Schlossgartenstra\ss{}e 9,
               D-64289 Darmstadt, Germany} 
\author{S.~Walter}
  \altaffiliation[Present address: ]{Institut
	    	f\"ur Kernphysik, FZ Karlsruhe,
		Postfach 3640, D-76021 Karlsruhe, Germany}
  \affiliation{Institut f\"ur Strahlenphysik,
               Universit\"at Stuttgart,
               Allmandring 3,
               D-70569 Stuttgart, Germany}
\author{J.~Enders}
  \email{enders@ikp.tu-darmstadt.de}
  \affiliation{Institut f\"ur Kernphysik,  
               Technische Universit\"at Darmstadt, 
               Schlossgartenstra\ss{}e 9,
               D-64289 Darmstadt, Germany} 
\author{H.~von~Garrel}
  \affiliation{Institut f\"ur Strahlenphysik,
               Universit\"at Stuttgart,
               Allmandring 3,
               D-70569 Stuttgart, Germany}
\author{O.~Karg}
  \altaffiliation[Present address: ]{Fachgebiet
		   Elektronische Materialeigenschaften,
		   Technische Universit\"at Darmstadt,
		   Petersenstra\ss{}e 23,
		   D-64287 Darmstadt, Germany}
  \affiliation{Institut f\"ur Kernphysik,  
               Technische Universit\"at Darmstadt, 
               Schlossgartenstra\ss{}e 9,
               D-64289 Darmstadt, Germany} 
\author{U.~Kneissl}
  \affiliation{Institut f\"ur Strahlenphysik,
               Universit\"at Stuttgart,
               Allmandring 3,
               D-70569 Stuttgart, Germany}
\author{C.~Kohstall}
  \altaffiliation[Present address: ]{Phywe Systeme,
		Robert-Bosch-Breite 10, D-37079 G\"ottingen, Germany}
  \affiliation{Institut f\"ur Strahlenphysik,
               Universit\"at Stuttgart,
               Allmandring 3,
               D-70569 Stuttgart, Germany}
\author{P.~von~Neumann-Cosel}
  \affiliation{Institut f\"ur Kernphysik,  
               Technische Universit\"at Darmstadt, 
               Schlossgartenstra\ss{}e 9,
               D-64289 Darmstadt, Germany} 
\author{H.\,H.~Pitz}
  \affiliation{Institut f\"ur Strahlenphysik,
               Universit\"at Stuttgart,
               Allmandring 3,
               D-70569 Stuttgart, Germany}
\author{V.\,Yu.~Ponomarev}
  \altaffiliation[Permanent address: ]{Joint Institute for Nuclear 
	       Research, 141980 Dubna, Moscow Region, Russia}
  \affiliation{Institut f\"ur Kernphysik,  
               Technische Universit\"at Darmstadt, 
               Schlossgartenstra\ss{}e 9,
               D-64289 Darmstadt, Germany} 
\author{M.~Scheck}
  \altaffiliation[Present address: ]{Department of
		Physics and Astronomy, University of Kentucky,
		Lexington, KY 40506-0055, USA}
  \affiliation{Institut f\"ur Strahlenphysik,
               Universit\"at Stuttgart,
               Allmandring 3,
               D-70569 Stuttgart, Germany}
\author{F.~Stedile}
  \affiliation{Institut f\"ur Strahlenphysik,
               Universit\"at Stuttgart,
               Allmandring 3,
               D-70569 Stuttgart, Germany}
\author{S.~Volz}
  \affiliation{Institut f\"ur Kernphysik,  
               Technische Universit\"at Darmstadt, 
               Schlossgartenstra\ss{}e 9,
               D-64289 Darmstadt, Germany}

\date{\today} 
 

\begin{abstract}
Results of a photon scattering experiment on $^{112}$Sn using bremsstrahlung 
with an endpoint energy of $E_0 = 3.8$ MeV are reported.  A $J = 1$ state
at $E_x = 3434(1)$ keV has been excited.  Its decay width into the ground state
amounts to $\Gamma_0 = 151(17)$ meV,
making it a candidate for a $[2^+ \otimes 3^-]1^-$ two-phonon state.  The results for
$^{112}$Sn are compared with quasiparticle-phonon model calculations as well
as the systematics of the lowest-lying $1^-$ states established in other even-mass 
tin isotopes.  Contrary to findings in the heavier stable even-mass Sn isotopes,
no $2^+$ states between 2 and 3.5 MeV excitation energy have been
detected in the present experiment.
\end{abstract}
\pacs{21.10.-k, 
      23.20.-g, 
      25.20.Dc, 
      27.60.+j  
	}
\maketitle 
 


Collective quadrupole and octupole vibrations are well-established features
of nuclear structure \cite{boh75}.  From the coupling of these elementary excitations
one expects multiplets of states whose energies and transition strengths
may provide insight into, e.\,g., the anharmonicities of vibrations, the
underlying microscopic structure of elementary excitations, or the
purity of multiphonon states.  For the specific case of the coupling 
of a quadrupole and an
octupole vibration, a quintuplet of states with
$J^\pi = 1^-$ to $5^-$ \cite{lip66,vog71} arises 
whose $1^-$ member is accessible by real photon
scattering \cite{kne96}.  Experimentally, the two-phonon structure of such $1^-$ states
has been established 
from the decay pattern in the case of some $N=82$ isotones \cite{wil96,wil98}.

A systematic study of two-phonon $1^-$ states using photon scattering
has been carried out for $^{116,118,120,122,124}$Sn 
by Bryssinck and co-workers \cite{bry99}.  Here, a surprisingly
uniform behavior concerning excitation energies and strengths has been identified,
in spite of a (slow) variation of the octupole collectivity with the mass number.
In addition, a number of $E2$ excitations between 2 and 4 MeV excitation energy
could be detected in the even-even $^{116-124}$Sn nuclei
in the photon scattering experiments \cite{bry00}.  Fair
agreement of a microscopic analysis with the experimental data has been found.


The present study aims at an extension of this survey to the most neutron-deficient
stable tin isotope, $^{112}$Sn, where
negative-parity states have been identified from 
inelastic neutron scattering \cite{kum05} recently, that are proposed to represent the members
of the quadrupole-octupole-coupled quintuplet.
Similar to the works of Bryssinck {\em et al.\/} 
\cite{bry99,bry00},
a photon scattering experiment has been performed at the nuclear resonance fluorescence
(NRF) setup at the Stuttgart Dynamitron accelerator \cite{kne96}.  Unpolarized 
bremsstrahlung was produced from a 3.8 MeV dc electron beam with average beam
current of 200 $\mu$A.  Photons scattered from a 1990-mg $^{112}$Sn target (enrichment
$\geq$ 99.5\%{}) have been measured in three high-purity germanium detectors
placed at 90$^\circ$, 127$^\circ$, and 150$^\circ$ with respect to the photon beam.
The detector at 127$^\circ$ was surrounded by BGO detectors for 
Compton background suppression.
Aluminum platelets \cite{pie95} and $^{13}$C powder \cite{mor93}
were placed around the Sn target 
for photon flux calibration.  Data were taken for 69 hours.

\begin{figure}
\epsfig{figure=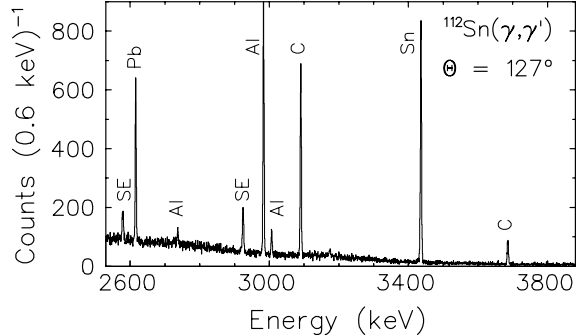,width=7.5cm}
\caption{\label{fig:spectrum}BGO-suppressed nuclear resonance fluorescence spectrum measured
at an angle of 127$^\circ$ with respect to the incoming beam direction.  A strong
dipole excitation at 3434 keV can be attributed to $^{112}$Sn while the other transitions
stem from $^{13}$C and $^{27}$Al used for the determination of the incoming photon spectrum.  
Single-escape peaks as well as a $^{208}$Pb background line are indicated.}
\end{figure}

Figure \ref{fig:spectrum} displays the measured spectrum for the example of the
detector placed at 127$^\circ$ with respect to the incident beam.  Besides 
transitions from the reference materials, $^{13}$C and $^{27}$Al, one recognizes
one strong transition at 3434(1) keV only to which dipole character 
can be unambiguously assigned on the basis of the measured angular distribution.  Aside from this
transition and from the decay of the $2_1^+$ state to the ground state (g.\,s.), 
no further transition has been
observed which could be attributed to $^{112}$Sn.  The sensitivity of the present experiment 
was comparable to the work of Bryssinck {\em et al.} \cite{bry99,bry00}.
A decay of the new 3434-keV state
into the $2_1^+$ state has also not been detected.  An upper limit of 1.5\%{} for the 
branching ratio of this decay can be extracted from the data.

In order to extract the excitation strength, the detector efficiency
and the bremsstrahlung photon spectrum need to be determined.  The relative shape of the
former was obtained from a measurement using a radioactive $^{56}$Co source, the
latter from the transitions in $^{13}$C and $^{27}$Al.  For a smooth interpolation
between the measured transitions, several assumptions about the shape of the photon 
distribution are possible:  (i) The shape of the photon spectrum may be simulated using
Monte-Carlo methods, (ii) it may be approximated by the Schiff function \cite{sch51} using the
endpoint energy of the spectrum as a free parameter, and (iii) it may be fit freely to
the measured transitions including the endpoint at which the photon intensity drops to zero.
One finds that neither approach can appropriately describe the 3088-keV transition from
$^{13}$C if one takes the latest literature value for the integrated cross section
determined from a self-absorption experiment \cite{mor93}.  
This behavior has been noticed before and is discussed in Ref.\ \cite{bec98}.
We have chosen to omit the 3088-keV line from the data points for the photon spectrum
determination and to use approach (ii) for the quantitative analysis as this resulted
in a very good description of the measured reference points.  Including the
lower $^{13}$C point will decrease the photon flux and hence increase the extracted
$B(E1)$ value by about 10\%{}.  
The shape of the photon spectrum as generated by a Monte-Carlo simulation 
using the code GEANT 3.21 does not
describe the data very well, in contrast to previous analyses (see, e.\,g., the work
by Belic {\em et al.\/} \cite{bel01}, where the simulation successfully described
the shape of the photon distribution).  


Table \ref{tab:results} lists the results for the observed dipole excitation in $^{112}$Sn
from the present work in comparison with the results of the other even-mass stable Sn
isotopes from Ref.\ \cite{bry99}.  Although the multipole character of the 3434-keV transition
and thus the parity of the $J=1$ state 
was not established in the experiment, we will henceforth assume that the transition
stems from the depopulation of the two-phonon $1^-$ state.  This is also suggested by the recent work
of Kumar and colleagues \cite{kum05}.  If a magnetic dipole character was assumed instead, the measured 
transition width into the g.\,s.\ of $\Gamma_0 = 151(17)$ meV 
would correspond to $B(M1)\uparrow = 0.97(11)~\mu_N^2$.  This would be
larger than in neighboring Cd \cite{koh05} and Te \cite{sch97,gul02} nuclei where
magnetic dipole strength was identified that might be attributed to the orbital $M1$ scissors mode
\cite{boh84,end05}.  No low-lying $M1$ strength was reported for the other even-mass tin isotopes
\cite{bry99,bry00}.

\begin{figure}
\epsfig{figure=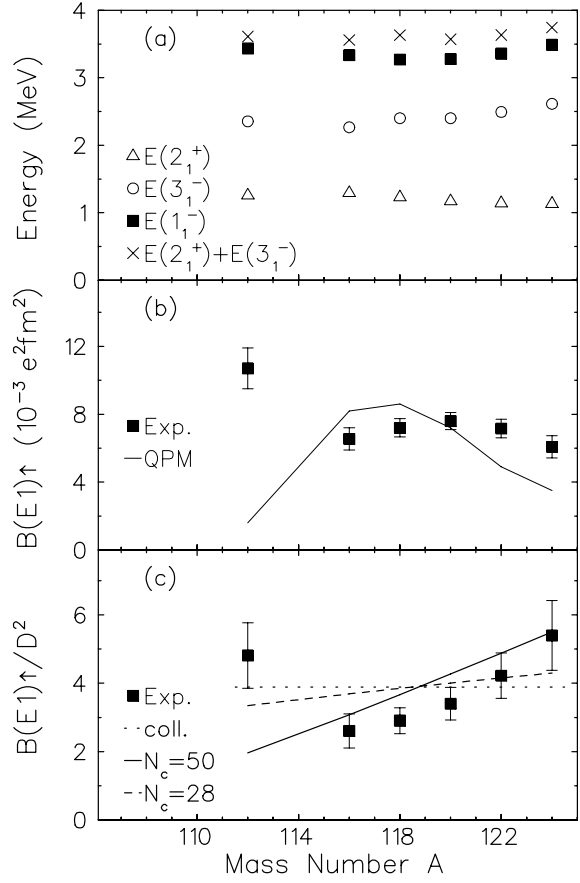,width=7.5cm}
\caption{\label{fig:results}Systematics of the $[2_1^+ \otimes 3_1^-]1^-$ two-phonon states in stable
even-mass tin isotopes, from this work and Ref.\ \protect\cite{bry99}.  (a) Excitation
energy of $1^-$ states compared with the excitation energies of the $2_1^+$ and $3_1^-$
states and their sums. (b) Measured $E1$ excitation strengths (full squares) compared
with the results of a quasiparticle-phonon model calculation (solid line).  (c)
Solid squares: Ratio of the measured $E1$ strength and the square of the dynamical dipole moment
according to Eq. (\protect\ref{eq:BMS}).  The lines
show the scaling behavior for a purely collective picture (dotted), for the assumption of
an $N_c = Z_c = 50$ core (solid), and for an $N_c = 28$ core (dashed).  The lines are
scaled to the mean of the experimental $B(E1)/D^2$ values.}
\end{figure}

From Table \ref{tab:results}
as well as from Fig.\ \ref{fig:results}(a), one recognizes that the excitation energies
of the low-lying $2_1^+$ and $3_1^-$ states vary only slowly with mass number, as does
the energy of the $1_1^-$ state.  The latter can be found, in all cases including $^{112}$Sn, 
about 5\%{} -- 8\%{} below the sum energy $E_x(2_1^+)+E_x(3_1^-)$.  

\begin{table*}
\caption{\label{tab:results}
Measured and calculated excitation energies and strengths for the lowest $1^-$ states
in $^{112,116,118,120,122,124}$Sn.}\begin{ruledtabular}
\begin{tabular}{lcccccc}
	& $^{112}$Sn\footnote{This work.} &$^{116}$Sn\footnote{Ref.\ \protect\cite{bry99}.} 
	&$^{118}$Sn\footnotemark[2] &$^{120}$Sn\footnotemark[2] 
	&$^{122}$Sn\footnotemark[2] &$^{124}$Sn\footnotemark[2]\\
\hline
$E_x(1^-)_{\rm exp}$ (keV) &
	3434 & 3334 & 3271 & 3279 & 3359 & 3490 \\
$E_x(1^-)_{\rm QPM}$ (keV) &
	3240 & 3350 & 3290 & 3320 & 3420 & 3570 \\
$E_x(2_1^+)+E_x(3_1^-)$ (keV) &
	3612 & 3560 & 3631 & 3572 & 3634 & 3646 \\
$B(E1)\uparrow_{\rm exp} \, (10^{-3}\,e^2$fm$^2$) &
	10.7(12) & 6.6(7) & 7.2(5) & 7.6(5) & 7.2(5) & 6.1(7) \\
$B(E1)\uparrow_{\rm QPM} \, (10^{-3}\,e^2$fm$^2$) &
	1.6   & 8.2 & 8.6 & 7.2 & 4.9 & 3.5 \\
\end{tabular}
\end{ruledtabular}
\end{table*}

In contrast to the very
uniform behavior of the excitation energies, 
the $E1$ excitation strength $B(E1)\uparrow = 10.7(12)\cdot 10^{-3}\,e^2$fm$^2$ 
of the $1_1^-$ state in $^{112}$Sn
is about 50\%{} larger than 
in the heavier even-mass Sn isotopes.  This result is depicted by the full squares in
Fig.\ \ref{fig:results}(b).  Our value from NRF is smaller than
the one reported by the (n,n$^\prime\gamma$) work \cite{kum05} which finds
$B(E1)\uparrow = (18^{+18}_{-5})\cdot 10^{-3}\,e^2$fm$^2$.  
Kumar and co-workers
were also able to determine a rough estimate for the $B(E1;3_1^- \rightarrow 2_1^+)$
value which is listed as $0.9(22)\cdot 10^{-3}\,e^2$fm$^2$.  The ratio
$B(E1;1^- \rightarrow 0_1^+)/B(E1;3_1^- \rightarrow 2_1^+)$ would be much larger
than expected from phenomenological \cite{pie99} and microscopic \cite{pon99} analyses,
but has a significant experimental uncertainty dominated by the measurement of the lifetime and
the multipole mixing ratio in the decay of the $3_1^-$ state.

To analyze the large $E1$ strength we used a phenomenological approach based on 
a collective dynamical dipole moment
as introduced by Bohr and Mottelson
\cite{boh57} and Strutinski \cite{str57}
\begin{equation}
	D_{\rm BMS} = 5.38\cdot 10^{-4}\cdot (Z+N) Z \beta_2\beta_3\,e\,{\rm fm},
	\label{eq:BMS}
\end{equation}
where the dipole moment arises from the dynamical quadrupole and octupole 
deformations $\beta_2$, $\beta_3$ as determined from the $B(E2;0_1^+\rightarrow2_1^+)$ and
$B(E3;0_1^+ \rightarrow 3_1^-)$ values, respectively, $Z$ denotes the proton and $N$ the
neutron number.  The ratio of the measured $B(E1)$ values and $D_{\rm BMS}^2$
is shown in Fig.\ \ref{fig:results}(c) and should be
constant within a collective approach (dotted line) in which the underlying shell structure
does not play a role.  While the heavier isotopes show a systematic increase with mass number,
the value for $^{112}$Sn is large and comparable to the result found for $^{122,124}$Sn.
Within the Bohr-Mottelson approach therefore, the uniformity of the $B(E1)$ values
in the heavier tin isotopes is surprising.  We note that in open-shell nuclei the ratio
$B(E1)/D^2$ is smaller by about an order of magnitude, as discussed by Refs.\ \cite{bab02,koh05}.

Recently, Kohstall {\em et al.} \cite{koh05} 
have applied a simple collective model description
to the two-phonon $E1$ excitations in the $Z\approx 50$ region.  Here, an effective dipole moment
is extracted
\begin{equation}
	D_{\rm eff} = \Delta e \frac{NZ}{A} \left( \frac{Z_v}{Z} - \frac{N_v}{N} \right),
	\label{eq:eff}
\end{equation}
where $Z_v$ and $N_v$ denote the number of valence protons and neutrons, respectively,
and $\Delta$ represents the distance between the centers of gravity of the proton and
neutron bodies.   Neglecting the shell structure, the dipole moment scales with
$e\Delta NZ/A$ so that a correction factor $K$ for an isotopic chain can be introduced to
account for shell effects
\begin{equation}
	[K(Z,N)]^2 = \left( \frac{N_c}{N} - \frac{Z_c}{Z} \right)^2\, ,
	\label{eq:K}
\end{equation}
which depends on the choice of the proton and neutron core, $Z_c = Z- Z_v$ and $N_c = N-N_v$,
respectively.
Scaled to the average, the correction factor $K^2$ 
can be compared to the experimental ratios
$B(E1)/D^2_{\rm BMS}$ as shown in Fig.\ \ref{fig:results}(c).  The full line indicates
the assumption of an $N_c =Z_c = 50$ core, the dashed line a core with $N_c = 28$ and $Z_c=50$.
While the heavier Sn isotopes nicely scale with the correction factor given by the
$N_c = Z_c = 50$ core, the value for $^{112}$Sn is found close to the assumption of an 
$N_c = 28$ core with the proton $Z_c = 50$ shell left intact, or it could be described 
by a collective
picture where the underlying shell structure does not play a role anymore.  However, 
although the experimental data are compatible with such an analysis, it is not
clear why the $N_c = 50$ shell closure should disappear when removing neutrons from a half-filled
$sdg$ neutron shell.  The unexpectedly large $E1$ strength in $^{112}$Sn thus remains unexplained.


In order to analyze the structure of the dipole excitation in $^{112}$Sn, we have also performed
calculations within the quasiparticle-phonon nuclear model (QPM \cite{sol76}) along the
lines of Refs.\ \cite{bry99,gri94,pon98}.  The results are displayed in Table \ref{tab:results}
with the experimental data for the $1_1^-$ states.  The calculated excitation strengths
are displayed as a solid line in Fig.\ \ref{fig:results}(b) and compared to the experimental
data.  The QPM fails to describe both the uniformity of the $E1$ strength found in
the heavier tin isotopes as well as the large $B(E1)$ value reported here for $^{112}$Sn.
The predicted $E1$ strength in this lightest naturally occurring tin isotope is very small.
Closer inspection reveals that the large contributions from protons and neutrons to the
transition matrix element connecting the ground with the two-phonon state nearly cancel each other.
The total transition matrix element is almost two orders of magnitude smaller than its
proton and neutron parts.
The calculation is thus very sensitive to the choice of parameters and does not exhibit too much
predictive power in this special case.


A few comments on the electric quadrupole strength distribution between 2 and 4 MeV 
are in order.  Several $2^+$ states are known in $^{116,118,120,122,124}$Sn. A
large fraction of the states below 4 MeV exhibits a strong
decay branch into the g.\,s.\ so that some of these states could be excited 
in the photon scattering study by Bryssinck {\em et al.\/} \cite{bry00}.
In the literature \cite{NDS112} for $^{112}$Sn, a number of 2$^+$ states are listed below 3.5 MeV.
The recent (n,n$^\prime\gamma$) work by the Kentucky group \cite{kum05} has found new $2^+$
states and corroborated previous $J^\pi = 2^+$ assignments, but different
branching ratios were reported in several cases.


The $E2$ excitation strengths and upper limits reported by Refs.\ \cite{kum05,NDS112} 
are consistent with the
upper limits derived from the detection threshold of the present experiment using the
branching ratios listed in the Nuclear Data Sheets \cite{NDS112}.
Summing the upper limits of all possible $E2$ excitations in the energy interval
between 2 and 3.5 MeV, one finds the maximum possible $E2$ strength to be $<260\,e^2$fm$^4$.
This is of the same order of magnitude as detected in the heavier Sn isotopes \cite{bry00}.
As no $E2$ transition could be detected in our experiment, it is likely that the $E2$
strength will continue the trend of the heavier isotopes to an even smaller summed $B(E2)$
value than in $^{116}$Sn which amounted to 143(19) $e^2$fm$^4$.
Calculations within the QPM expect nearly a dozen $2^+$ states between the $2_1^+$ state and
3.8 MeV, none of which is predicted to be strong enough to
be detectable in the present experiment.  


In conclusion, we have performed a nuclear resonance fluorescence experiment using
bremsstrahlung with an endpoint energy of 3.8 MeV on the semi-magic $^{112}$Sn nucleus.
The strength of a dipole excitation at 3434 keV has been determined, and this excitation
is assumed to arise from the coupling of the low-lying quadrupole and octupole vibrations.
Compared with the heavier even-mass stable tin isotopes, the $E1$ strength of the
3434-keV excitation in $^{112}$Sn is about 50\%{} larger.  
We have tried a phenomenological analysis which does not lead to a
qualitative understanding of the large $B(E1)$ value in $^{112}$Sn.
The detected $E1$ strength cannot be reproduced by a microscopic quasiparticle-phonon
model calculation that turns out to be very sensitive to the interplay of proton and
neutron amplitudes in this transition.
No experimental evidence for $E2$ strength above the
$2_1^+$ state was found.
Measuring the form factors of the low-lying
$1^-$ and $2^+$ states in electron scattering could provide a quantitative estimate of 
the composition of the wave function and admixtures from collective excitations.
To elucidate the change of the $E1$ strengths of the $1^-$ two-phonon states 
in $^{112}$Sn and $^{116}$Sn, it would be desirable to obtain information about the
only missing stable even-mass tin isotope, the rare $^{114}$Sn.  Also, the multipole
character of the 3434-keV excitation should be investigated.

\begin{acknowledgments} 
The support of the operations staff at the Dynamitron accelerator in Stuttgart
is gratefully acknowledged.  We are indebted to the target laboratory of the
Gesellschaft f\"ur Schwer\-io\-nen\-for\-schung (GSI) for the loan of the target material
and to A.\ Richter and A.\ Zilges for helpful discussions.
This work is supported by the Deutsche Forschungsgemeinschaft through
Sonderforschungsbereich 634 and through contract Kn 154-31.
\end{acknowledgments}

\end{document}